\documentstyle[epsfig]{elsart}
\newcommand{\bea}{\begin{eqnarray}}
\newcommand{\beq}{\begin{equation}}
\newcommand{\eea}{\end{eqnarray}}
\newcommand{\eeq}{\end{equation}}

\begin{document}
\runauthor{L. Benet and T.H. Seligman}

\begin{frontmatter}
\title{ Generic occurrence of rings in rotating systems }
\author{ L. Benet\thanksref{LBF}} and 
\author{ T. H. Seligman }
\address{
Centro de Ciencias F\'{\i}sicas, University of M\'exico (UNAM), 
Cuernavaca, M\'exico\\
and\\
Centro Internacional de Ciencias, 
Cuernavaca, M\'exico}

\thanks[LBF]{ Present address: Max-Planck-Institut f\"ur Kernphysik,
D-69117 Heidelberg, Germany. E-mail: Luis.Benet@mpi-hd.mpg.de}

\date{\today}

\begin{abstract}
In rotating scattering systems, the generic saddle-center scenario 
leads to stable islands in phase space. 
Non-interacting particles whose initial conditions are defined in
such islands will be trapped and form rotating rings.
This result is generic and also holds for systems quite different 
from planetary rings. 
\end{abstract}

\end{frontmatter}

Light particles interacting with massive rotating systems are 
frequently encountered in diverse fields of physics. 
Electrons in rotating molecules and particular versions of the 
restricted three-body problem in celestial mechanics are the most 
well-known examples, but some nuclear models 
fall into the same category. 
Situations with large angular momenta will be of particular interest.
Increasing amounts of information about narrow planetary rings suggest 
that such rings are often associated with the so-called shepherd 
satellites~\cite{GT79}, and may exist due to mechanisms somewhat 
more complicated than the well known broad 
rings~\cite{planetaryrings,ringS}.
The implications of such mechanisms for the above mentioned systems 
could be far reaching. 

The purpose of this paper is to show that there exists in rotating 
systems a generic mechanism to obtain narrow rings with structure, 
that does {\it not} depend on Kepler orbits. 
The genericity of the mechanism guarantees that the stable orbits 
supporting the ring structure are rather insensitive to small 
perturbations and thus may play a role in different situations of 
the type mentioned above. 

The principal result of the paper is a very general argument 
to support the occurrence of such rings in terms of one of the two 
generic scenarios for the formation of bounded trajectories in a 
scattering problem as a function of some external parameter. 
We exemplify this argument using the simple model of discs 
rotating around a center outside of these discs, which we proposed 
earlier~\cite{meyeretal95}. 
We shall find a large number of sometimes complicated rings.
The introduction of a second disc on a smaller orbit 
may limit these to a small number of narrow rings.
Our line of argumentation does not require that both discs move 
at the same angular velocity to maintain a ring structure.
This implies that we leave the framework of a rotating system and with 
it the conservation of the Jacobi integral; the generic properties 
play an important role in the transference of the results obtained to
situations for which the Jacobi integral is not conserved.

If we transform free motion to a synodic frame (rotating frame) the 
Jacobi integral becomes the Hamiltonian and the motion becomes a 
two-armed spiral.
One arm for the incoming motion up to the point of closest approach 
to the center of rotation, and another for the outgoing motion after 
this point. 
We next consider a convex repulsive potential rotating at some 
distance from the center. 
It is clear that this potential can throw a particle from the outgoing 
arm of one spiral to the incoming arm of another.
If the radial motion of the particle is sufficiently slow that the 
particle can hit the same repulsive potential again on its new way 
out, the trajectory may be confined.
This will occur if the absolute value of the Jacobi integral is 
sufficiently small to avoid that the particle will always 
escape after at most one collision.
Thus by gradually reducing (or increasing for negative values) the 
Jacobi integral we will find bounded orbits at some value. 
There are two generic ways for this to occur: First, a saddle-center 
bifurcation which will always produce a stable and an unstable 
periodic orbit (elliptic and hyperbolic fixed points)~\cite{DGOY90}. 
Second, the scenario where abruptly a fully hyperbolic structure 
appears in phase space~\cite{TGO93}; the latter are 
typically associated with maxima in the potential and what is known 
as orbiting scattering.

\begin{figure}
\noindent\centerline{
\psfig{figure=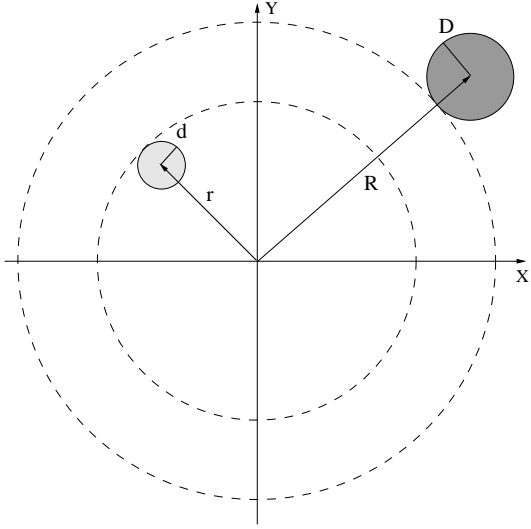,width=7cm,angle=0}}
\caption{
\label{fig1} 
Geometry of the two rotating discs scattering billiard: $R$ and $r$
define the radial position of the discs; $D$ and $d$ their radii. 
Both discs rotate about the origin with frequencies $\Omega$
and $\omega$.}
\end{figure}

We shall concentrate on the first scenario, for which the occurrence 
of stable periodic orbits is generic. 
In the synodic frame non-interacting particles dispersed along the 
corresponding stable island will usually form an eccentric ring 
obtained from the spiral orbit deformed by the potential such as to 
form a closed path. 
In the sidereal (space-fixed) frame this ring will undergo a precession 
corresponding to the frequency of the rotation. 
As we shall see below, the ring obtained in the sidereal frame may be
quite different from the actual trajectories of the individual ring 
particles.

To illustrate the general argument we shall start from a toy-model, which 
has been studied before~\cite{meyeretal95}.
The repulsive potentials in this case corresponds are two (rotating) 
hard-wall scatterers.
This model has the additional advantage of excluding the second 
generic scenario mentioned above.

We consider first one disc rotating around a center which lies outside 
this disc (Fig.~\ref{fig1}); the motion will be restricted to a plane.
We shall denote by $R$ the radial position of the center of the disc 
with respect to the center of rotation, its radius by $D$ ($R>D$) and 
the angular velocity by $\Omega$.
In the sidereal frame, particles move on straight lines with constant
velocity until they collide with the disc; otherwise they escape.
A collision with the disc will typically change the direction and
the magnitude of the velocity of the particle.
Only if the collision is radial, i.e. in one of the intersection 
points on the disc of the line that joins the center of rotation
(origin) with the center of the disc, the magnitude of the velocity 
will be unchanged, and the incoming collision angle is equal to the
outgoing one.
For these orbits, we can choose the outgoing angle $\alpha$ and the 
velocity of the particle to obtain identical and consecutive radial 
collisions.
In the synodic frame, these orbits are periodic and symmetric.
Such orbits provide the backbone of the horseshoe 
construction~\cite{benetetal99}, as was shown for this model in 
Ref.~\cite{meyeretal95}.
In terms of the Jacobi integral, these orbits are given by
\beq
J_n=2\Omega^2 (R-D)^2
{ \cos^2\alpha-[(2n+1){\pi\over 2} -\alpha]\sin 2 \alpha
\over [(2n+1)\pi - 2 \alpha ]^2}.
\label{sympo}
\eeq
Here, $\alpha$ is measured with respect to 
the normal at the radial collision point, and $ n $ is the number of 
complete rotations the disc performs between consecutive collisions.

\begin{figure}
\noindent\centerline{
\psfig{figure=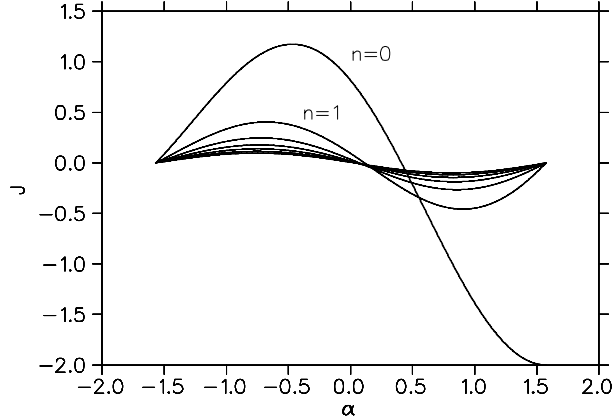,width=5.5cm,angle=90}}
\caption{
\label{fig2} 
Chart of symmetric periodic orbits of the rotating disc billiard 
for various values of $n$ ($R=3$, $D=1$ and $\Omega=1$).}
\end{figure}

In Fig.~\ref{fig2} we show for small $n$ the characteristic 
curves $J_n$ for the symmetric periodic orbits.
As mentioned above, there is a connected interval of the Jacobi 
integral where the action of the repulsive potential can build 
periodic orbits.
In fact, by reducing the absolute value of the Jacobi integral, a 
saddle-center bifurcation creates a pair of periodic orbits every time
a maximum or minimum is crossed. One of these is stable and the other 
unstable~\cite{meyeretal95}.
In the neighborhood of every stable periodic orbit there exist small
regions of stability, which allow for the appearance of rings 
as described above.

\begin{figure}
\noindent
\centerline{\psfig{figure=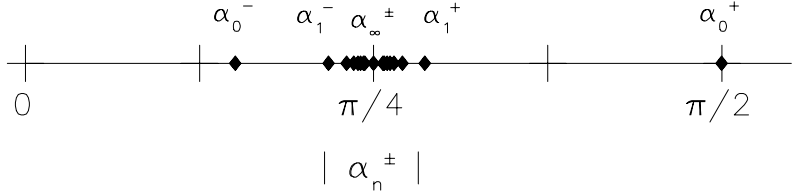,width=2.5cm,angle=90}}
\caption{
\label{fig3} 
Absolute value of the solutions of Eq.~(\ref{trascendent}), giving 
the angles $\alpha_n^\pm$ where new periodic orbits are created by 
saddle-center bifurcations.
The stable periodic orbits will appear for values slightly to the 
right of the points shown.}
\end{figure}

As the stability of the symmetric periodic orbits is 
known~\cite{dullin}, we can actually determine at what angle $\alpha$ 
stable orbits, periodic in the synodic frame, can occur. 
Indeed, for each $n$ there is exactly one prograde ($\alpha>0$) and one 
retrograde orbit ($\alpha<0$), that will be stable over some interval 
of decreasing $\vert J \vert$ and then undergo a period doubling 
bifurcation sequence. 
The only exception is the prograde $ n=0 $ solution which is marginally 
stable for a single $ J $ at $ \alpha = \pi /2 $.
The angles $ \alpha _n^\pm $ for which each saddle-center bifurcation 
occurs are given by
\beq
\tan\alpha^\pm_n={2\over{(2n+1)\pi-2\alpha^\pm_n}}\pm 1.
\label{trascendent}
\eeq
Here, the upper index identifies the sign of the solution of 
Eq.~(\ref{trascendent}), that is whether the solution corresponds to
a prograde or retrograde trajectory.
The absolute value of the solutions of this equation are shown in 
Fig.~\ref{fig3}.
The stable periodic orbits are found at absolute values of
angles slightly larger than these~\cite{dullin}.

\begin{figure}
\noindent
\centerline{\psfig{figure=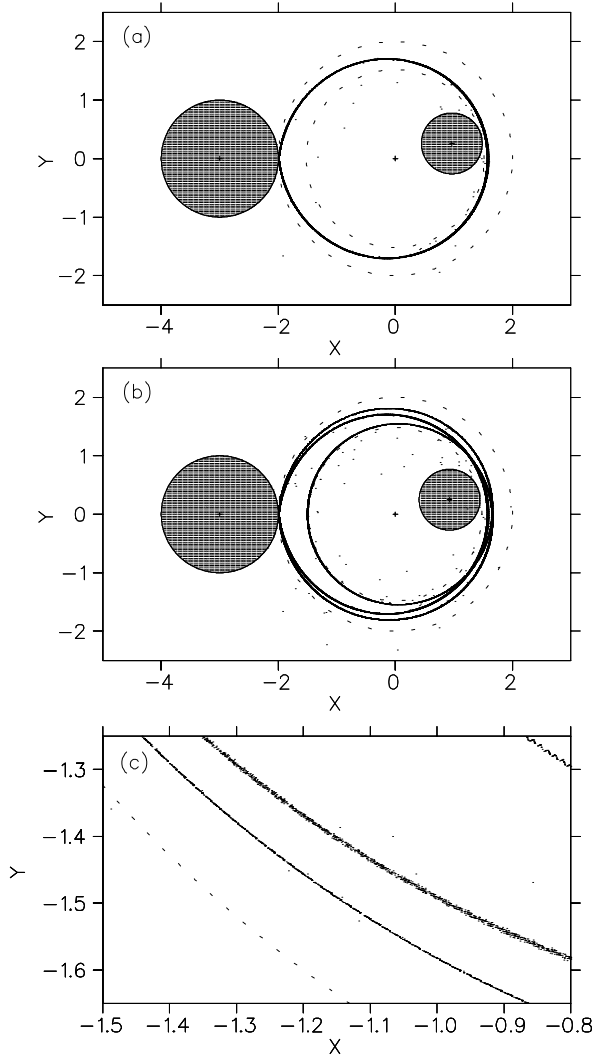,width=14cm,angle=90}}
\caption{ 
\label{figs4} 
Rings of the two-disc billiard in the sidereal frame ($R$ and $D$ 
chosen as in Fig.~\ref{fig2}).
(a)~Only the $\alpha_1^+$ component survives ($r=1$, $d=0.52$);
(b)~the $\alpha_2^+$ component is also present ($r=0.96$, $d=0.52$).
(c)~Detail of~(b) showing the finite width of the ring.}
\end{figure}

Ring structures are obtained if we distribute initial conditions 
randomly in the interaction region and observe those that remain 
after a long time (Figs.~\ref{figs4}).
Here, we have assumed that ring particles do not interact among 
themselves.
If we consider all these stability regions together, we find a rather 
large area of rings that can coexist if there is no restriction on 
the initial conditions.
This would lead to a wide ring with very complicated structure.
Each stable region contributes a narrow ring that may have several 
loops, giving rise to different strands.

One way to obtain well defined narrow rings with only their intrinsic 
structure would be to limit initial conditions to the surroundings of 
one of the stable islands (as we often do for numerical purposes in 
our Monte Carlo calculations). 
Yet we do not want to argue such a selection, as this would imply 
information about the formation of rings, which is by no means the 
subject of this Letter. 
Based on the known presence of two shepherds near some narrow 
rings~\cite{GT79}, we introduce a second disc moving 
on a circular trajectory with respect to the same center that lies 
inside the one we have considered above (see Fig.~\ref{fig1}). 
We shall proceed to show that a second disc indeed provides such a 
selection mechanism.

The main effect of a second disc on a smaller orbit will be to sweep 
many of the possible stable orbits. 
The corresponding elliptic regions and therefore the associated
rings will disappear.
Clearly, new ones will be created inside the inner edge of the inner 
disc, but those are again of the type just discussed and we shall 
disregard them.  
Also, new periodic orbits involving collisions with both discs will 
show up, but these tend to be very unstable. We shall thus concentrate 
on the periodic orbits of the outer disc that are not affected by the 
inner one.

The  simplest case is, in some sense, the one where the two discs move 
with incommensurable frequencies, although this implies that the 
Jacobi integral is no longer conserved.
We proceed to evaluate which periodic orbits will not be affected 
by the inner disc, whose center is at distance $r$ from the rotation 
point and whose radius is $d$.
We assume that the orbits of the discs are non-overlapping, i.e.,
$r+d<R-D$.
From the geometrical arrangement, it is clear that all the orbits 
that cross the outer edge of the inner disc will be affected by this 
disc.
In terms of the angle $\alpha$, we find that the orbits that will not
be affected satisfy 
\beq
|\alpha|\ge \alpha_{\rm max} = \arcsin {r+d\over R-D}.
\label{alfmax}
\eeq
Note that for commensurable and in particular equal frequencies, 
resonant conditions may preserve some ring components for special 
configurations, while for other configurations these components may 
be wiped out. 
Yet, for the survival of rings fulfilling Eq.~(\ref{alfmax}), no 
resonance condition is necessary.

Equation~(\ref{alfmax}) permits to predict, which components will 
be unaffected by the inner disc.
In this sense, Eq.~(\ref{alfmax}) and the precise positions of
$\alpha^\pm_n$ define selection rules.
For instance, if the geometry is such that the condition 
$\alpha_2^+<\alpha_{\rm max}<\alpha_1^+$ holds, the system will 
only display one ring corresponding to the $\alpha_1^+$ stable 
region (see Fig.~\ref{fig3}). 
For $\alpha_{\rm max}$ sufficiently larger than 
$\alpha_1^+$, rings will not occur in the system, while for 
$\alpha_{\rm max}<|\alpha_0^-|$ all possible strands will show up.
We conclude that ring components corresponding to prograde 
orbits are more likely to be found than those associated to retrograde 
orbits.

\begin{figure}
\noindent
\centerline{\psfig{figure=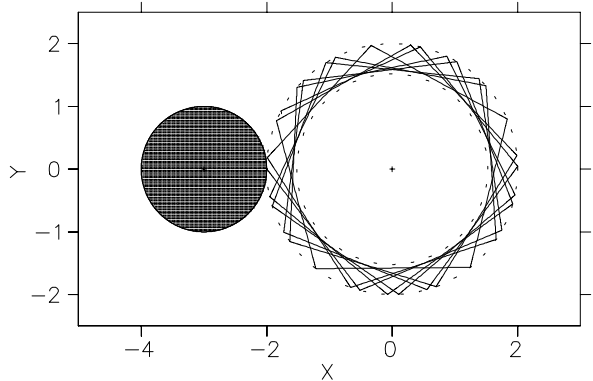,width=5cm,angle=90}}
\caption{ 
\label{fig5} 
Trajectory in the sidereal frame of a ring particle corresponding to 
$\alpha^+_1$. 
The initial conditions were set in the position where the disc is 
shown.
The dotted circles represent the paths drawn by the inner and outer 
edges of the outer and inner discs respectively.}
\end{figure}

In Fig.~\ref{figs4}a we present in the sidereal frame, 
examples of the ring structures found when only the $\alpha_1^+$ 
component survives, and in Fig.~\ref{figs4}b
the case when the $\alpha_2^+$ ring is also present.
Figure~\ref{figs4}c shows the finite width of the rings in 
an amplified region. 
As a function of time, these rings rotate with the frequency of the 
outer disc. 
While these figures show rings at fixed time in the sidereal frame, 
the individual particles follow polygonal orbits corresponding to 
reflection angles near $a^+_n$, as shown in Fig.~\ref{fig5}. 
These polygons will typically not close.
Note that the corresponding periodic orbits in the synodic frame will
have a shape similar to the rings in the sidereal frame. 

If we would use a mountain-like potential rather than a hard disc, 
associated with the hilltop the second type of scenario, an abrupt 
bifurcation, may occur~\cite{TGO93}.
This scenario implies the sudden appearance of a hyperbolic structure
which is structurally stable.
But in these case pruning would typically set in after a finite
change of the Jacobi integral and we revert to the other scenario 
if the mountain has a steep slope.

For attractive potentials we can have other periodic orbits, but 
typically we would expect that the ones we consider still exist, and 
follow a similar scenario.
The case of attractive gravitational potentials is of particular 
interest: the fact that the central potential will produce elliptic 
or hyperbolic trajectories in the sidereal system can easily be 
included in the argument.
On the other hand, the existence of the weaker $ 1/r $ potentials from 
the shepherd moons seems to fit only loosely in the picture discussed 
above. 
Yet, we may recall that H\'enon~\cite{henon} found that the periodic 
orbits of consecutive collision in the restricted three-body problem 
for zero mass parameter (the Kepler problem with a non-attractive 
rotating singularity) determine the structure of the periodic orbits 
for finite small masses. 
This obviously carries over to the four-body problem. 
Therefore, we can expect the generic hard-disc results to have 
some qualitative similarity to the shepherd situation, except 
that the roles of the inner and outer shepherds may be
interchanged~\cite{ringmar}.
This argument is further supported by some numerical investigations
in the restricted three-body problem for small mass ratio~\cite{celmec2},
as well as by the recent rigorous proof, for the same problem, of the 
existence of a chaotic subshift near collision orbits ~\cite{BMacKay00}.
Further research in this direction is on the way.

Finally we would like briefly to touch upon the structure of the narrow 
rings we found.
As mentioned above and illustrated in Figs.~\ref{figs4}, stable 
periodic orbits belonging to different stability regions lead to 
independent ring components, each of which may display several loops 
($n+1$ for retrograde and $n$ for prograde orbits). 
Particles in different rings will clearly have different speeds.
There is a second mechanism that does not imply such a 
difference of speeds.
A single strand will generically become structured, when the elliptic 
fixed point undergoes the period doubling cascade.
Just after the period doubling, the ring component will have first two,
then four, etc., entangled strands associated with each region of 
stability.
This mechanism can produce a narrow braided ring.
In this case ring particles in the different strands move 
almost synchronously.
Indeed, the relative motion of ring particles and system rotation 
is a result of the particular potential and may be near zero for the 
first stable prograde orbit. 

In conclusion, we have established the existence of a generic scenario 
for the occurrence of stable rings of non-interacting particles in 
rotating systems. 

The example we discussed above is particular in several senses:
First, we exclude the abrupt bifurcation scenario. 
If instead of hard discs we consider smooth potential hills of a 
Gaussian or similar shape, this scenario will occur at the hill tops, 
and give rise to interesting phenomena. 
Yet in their steep flanks the usual saddle-center bifurcations will 
still take place.
Second, we violate the invariance of the Jacobi Integral only in a 
subspace of phase space, that we are not interested in. 
If we think of semi-classical applications, the border of the 
invariant subspace will be smeared out and the symmetry breaking 
will become ubiquitous though it may be weak in some parts.
It is here that the genericity of saddle-center bifurcations and their 
consequent insensitivity becomes very important, as it guarantees that
the structures survive with minor changes.
Third, the rings, as found here, display complicated structure, both 
because two or several rings with different particle speed may coexist
and interfere in complicated ways in semi-classics, and because a 
single ring may have strands of similar particle speed as the central 
orbit undergoes a period-doubling cascade.

In semi-classics we can proceed to perform the calculation in the 
rotating system, where we will use standard techniques both for 
the stable and unstable orbits; thereafter, we can transform the 
resulting wave function to space fixed system.
The errors of the semi-classical approximation will be modified, but 
the result in the space fixed frame will be correct to the same order 
in $\hbar$. Thus we can expect to see such 
structures whenever the motion of the light particle is too slow for 
the Born-Oppenheimer approximation to be valid, but semi-classical 
reasoning is adequate. This will {\it e.g.} be the case for very 
large angular momenta both in molecules and Nuclei.

In this letter we emphasize the generic character of the result 
obtained.
The possible applications have to be studied within each particular 
field.
In most cases the fact, that we expect the same behavior for 
attractive forces is very important. In nuclei, for example, the 
rotating mean potential may well result in resonances of surface 
nucleons that spend most of the time outside this potential. 
In molecules high angular momenta for the core are still quite 
inaccessible to most experiments, but it will be interesting to see 
what happens at the verge of the formation of Rydberg states in such 
cases.
Finally we should not discard the possibility that such effects are 
relevant to argue that narrow rings with shepherds may live longer 
than the broad rings due to the generic stability of such 
structures~\cite{ringmar}.

We are thankful to Fran\c{c}ois Leyvraz and Christof Jung for many
useful discussions.
We also acknowledge one of the referees for bringing to our attention 
Ref.~\cite{BMacKay00}, which gives a stronger support to our ideas.
This work was partially supported by the DGAPA (UNAM) project IN-102597
and the CONACYT grant 25192-E.

\end{document}